\documentclass[pre,twocolumn,nofootinbib]{revtex4-1}
\usepackage{amsmath}
\usepackage{bm}
\usepackage{amssymb}
\usepackage{tikz}
\usetikzlibrary{cd}
\usepackage{mathtools}

\usepackage{multirow}
\usepackage{adjustbox}
\usepackage{bbm}
\usepackage{tabularx}
\usepackage{graphicx}
\usepackage{xcolor}
\usepackage{ulem}
\usepackage{ifthen} 

\begin{document}

\title{A Bouquet for Apollonius: Focal Conics in Sessile Cholesteric Droplets}

\author{Randall D. Kamien$^a$}
\author{Yuriy Nastishin$^b$}
\author{Brigitte Pansu$^c$} \email{ brigitte.pansu@universite-paris-saclay.fr}

\affiliation{$^a$Department of Physics and Astronomy, University of Pennsylvania,
209 South 33rd Street, Philadelphia, Pennsylvania 19104, USA}
\affiliation{$^b$Hetman Petro Sahaidachnyi National Army Academy, Lviv, Ukraine}
\affiliation{$^c$Universit\'e Paris-Saclay, CNRS, Laboratoire de Physique des Solides, UMR-8502, 91405, Orsay, France}

\begin{abstract}
 Focal conic domains, are defects characteristic of layered liquid crystal phases. Their association can built flowers where petals are the ellipses of the Dupin cyclides involved in these defect.  We report here the observation of focal conic flowers in cholesteric droplets sessile on a glass surface and surrounded by glycerol. The observation of the droplets in different directions helps to solve the 3D architecture of the flower. The effects of the droplet size and of the pitch value are also reported. 
\end{abstract}

\maketitle

\vspace{2 mm}

{T}he accidental discovery of liquid crystals was aided by the presence of optically striking patterns.  In particular, the cholesteric phase, first observed by Planer  \cite{Planer1861,NAZARENKO201829} and later by Reinitzer \cite{Reinitzer} provide enigmatic and beautiful textures arising from the spontaneous yet perfect arrangement of molecules at the sub-micron scale; the cholesteric phases that they observed are comprised of highly anisotropic molecules with their long axes lying perpendicular to and rotating periodically about the helical axis on a length scale close to the optical range.  As a result the cholesteric acts as a diffraction grating \cite{dgp}, affording the observer a labradorescent splendor.  While these discoveries were being made, Maxwell developed the theory of canal surfaces; surfaces swept out by a sphere of varying radius moving along an arbitrary path.  These surfaces are everywhere normal to the rays emitted perpendicular to the path.  Among the most intriguing are the cyclides of Dupin \cite{Dupin} which are canal surfaces that are generated by two distinct paths and it was established that the two paths were an ellipse and a single branch of a hyperbola in the plane perpendicular to the ellipse. Moreover, the ellipse passes through the focus of the hyperbola just as the hyperbola passes through the focus of the ellipse.  Later, it was realized by Friedel and Grandjean \cite{Friedel} that these ``confocal'' conic sections were present in smectic liquid crystals, materials that, like the cholesteric, enjoy a one-dimensional modulation, in density or orientation for the smectic or cholesteric, respectively, {such, that liquid crystal layers adopt the shape of cyclides of Dupin, which are canal surfaces, the rays perpendicular to which correspond to the nematic director in the smectic and to the helical twist axis in the cholesteric.}  
\begin{figure}[t]
 \centering
  \includegraphics[width=0.4\textwidth ]{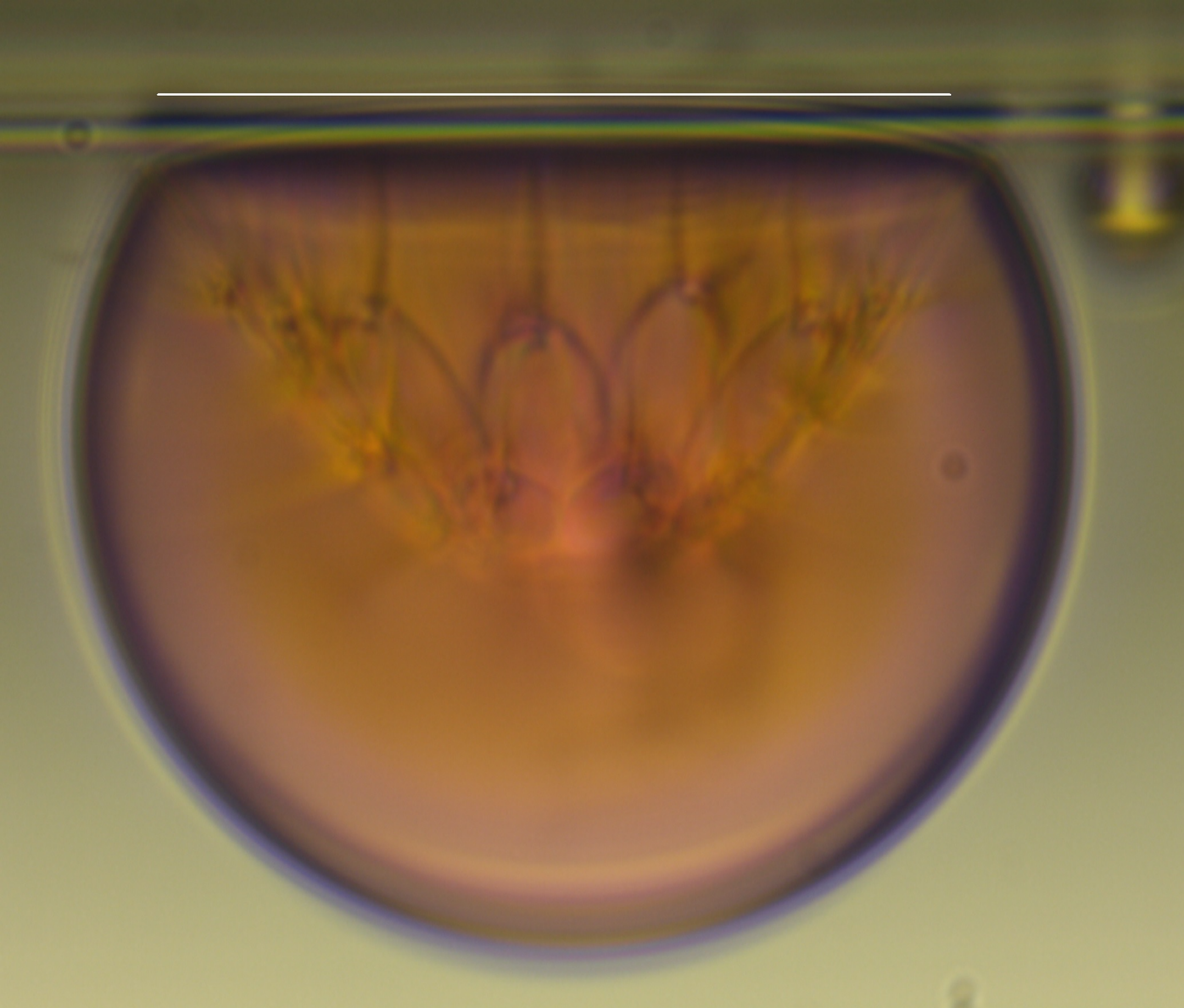} 
 \caption{ \label{fig:Droplet24pc_2} A sessile cholesteric droplet that is stuck to a flat glass capillary (bright field).  This is a view from the side and shows a clear interface inside the droplet.  It is a paraboloid.  The paraboloid is further decorated with ellipses that indicate the presence of FCDs. {Because of optical distortion only limited parts of the hyperbol\ae\ are visible. Varying the microscope focus  can reveal parts of hyperbol\ae\ that were not clear originally.  For this reason, we show here only illustrative photographs.} Scalebar 100$\mu$m.}  \end{figure}

\begin{figure}
 \centering
\begin{tabular}{ c c  }
\begin{minipage}{0.29\textwidth}
  \includegraphics[width=\textwidth]{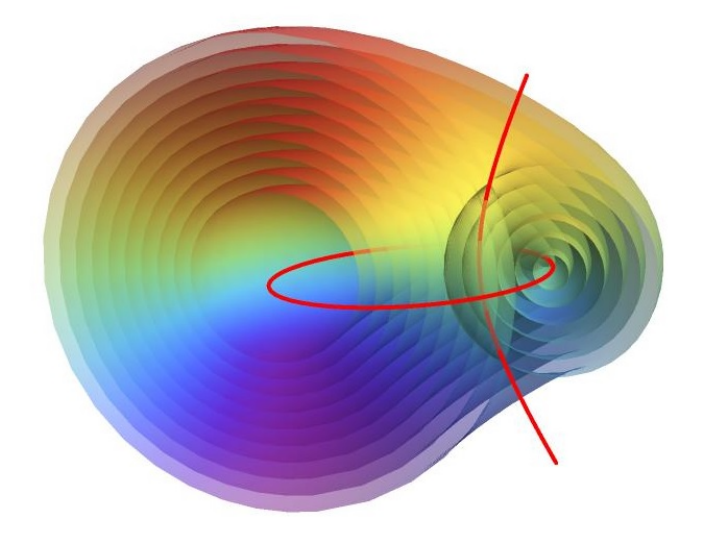}
  \end{minipage}&
  \begin{minipage}{0.16\textwidth}
 \includegraphics[width=\textwidth]{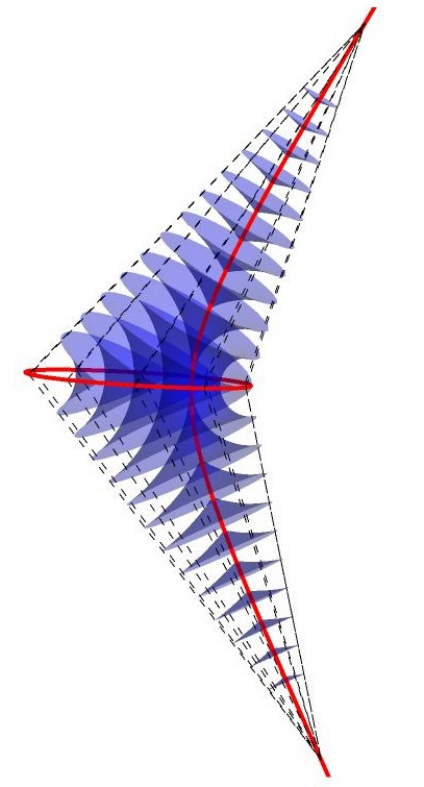}
 \end{minipage}\\(A)&(B)\\
 \end{tabular}
 \caption{ \label{fig:CyclideDupin} A) The Dupin cyclides   \cite{Dupin} are canal equidistant surfaces that are generated by  an ellipse and a single branch of a hyperbola in the plane perpendicular to the ellipse.  Moreover, the ellipse passes through the focus of the hyperbola just as the hyperbola passes through the focus of the ellipse B) Example of a  focal conic domain (FCD).  A classical focal conic domain is part of the Dupin cylides limited by a cone resting on the ellipse and with its top located on the hyperbola.The angle $\gamma$ between the two asymptotes of the hyperbola is 120$^o$.}
\end{figure}

Here, we report on a cholesteric droplet shown in Fig. \ref{fig:Droplet24pc_2} seated on a glass substrate that forms a complex bouquet of domains filled with sections of the cyclides of Dupin following the precise rules of focal conic domains (FCDs).  Visible in optical microscopy, this demonstrates the ability to control structure through surface cues in addition to providing a remarkably clean and precise display of geometry in a cholesteric system.  {Recall that Dupin cyclides are equidistant curved layers, segments of equidistantly nested surfaces that, like the torus, have circles as their lines of curvature (Fig. \ref{fig:CyclideDupin}A). In the smectic  and cholesteric, {only segments of single sign (usually negative) Gaussian curvature} have been reported to form in a single FCD. In an FCD, the long axes of the molecules are along straight lines connecting the ellipse and real branch of hyperbola; these straight lines are generatrices of cones whose apices are on the hyperbola and their common base is the ellipse (Fig. \ref{fig:CyclideDupin}B).}


 
Typically, FCDs are studied in the smectic {A} liquid crystal phase.  There, the one-dimensional density wave of the smectic can be modeled via a Canham-Helfrich like energy \cite{Canham,Helfrich}, embellished by a compression modulus maintaining a fixed spacing between otherwise featureless layers.  The energetics of these materials have been widely studied \cite{deGennes1972,AvilesGiga,bps} and, in particular, the energetics of the FCDs have been analyzed \cite{Fournier,Kleman}.  However, if one chooses to build complex layered structures with the FCDs as geometrically perfect building blocks, then they can be attached following purely geometric rules \cite{Friedel,Bouligand,KlemanSethna,Kleman2000,BellerFlower}.   The analysis of our observations will come in two steps.  First, we will treat the cholesteric as if it were made of layers (as in the smectic) and deconstruct our observed images in terms of focal conic domains.  We will then go back and decorate the smectic layers with director fields, reintroducing cholesteric structure.  It has been shown that this construction is possible when the original smectic layers are constructed from FCDs.  Since the cholesteric does not, {\sl per se}, have layers, we will call these surfaces pseudolayers in the following.


 \begin{figure}[b]
 \centering
 \includegraphics[width=0.45\textwidth]{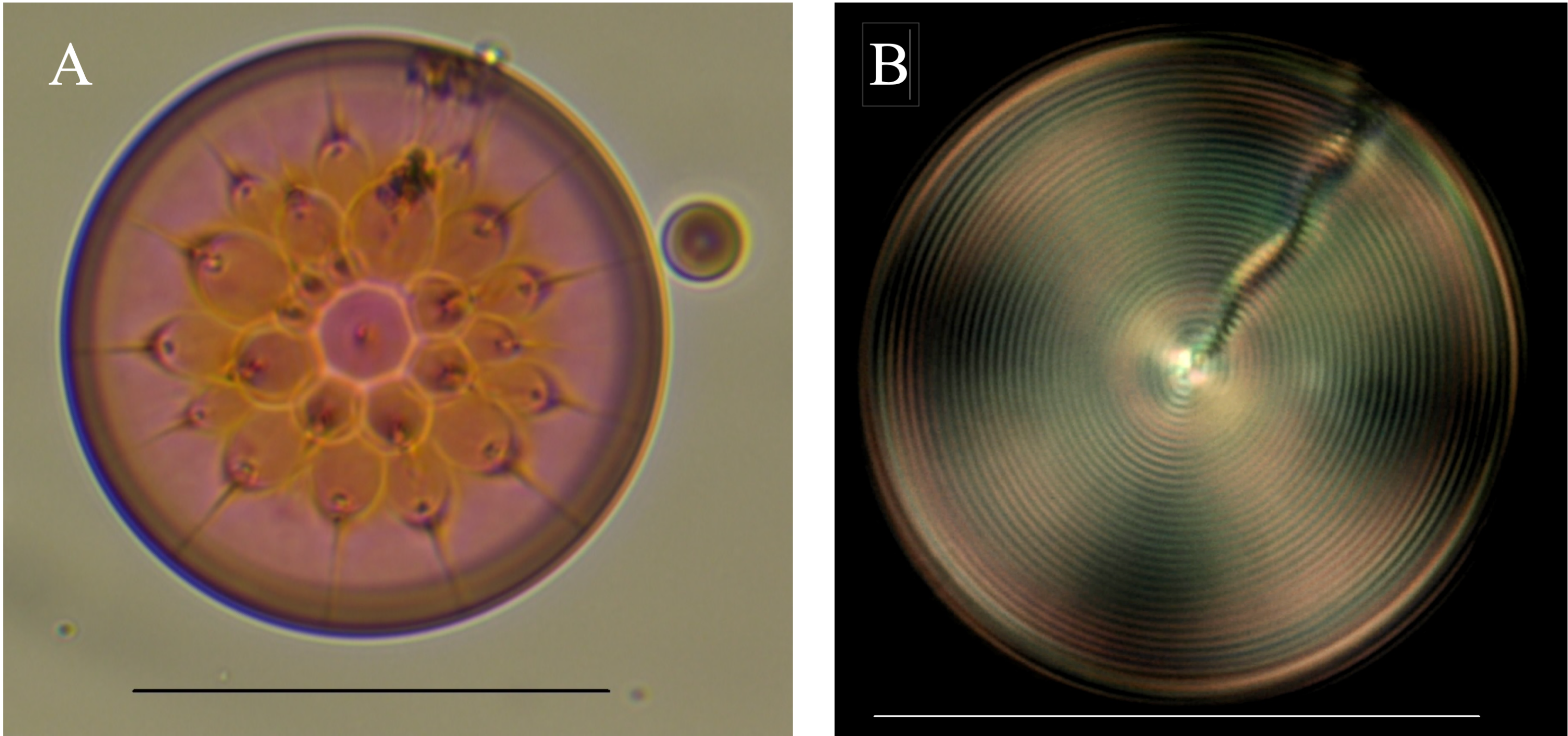}
 \caption{ \label{fig:Droplet24pc} Cholesteric droplets in glycerol.  A) Sessile, short-pitch (0.33$\mu$m) cholesteric droplet seen from below in bright field without any polarizer.  Note the presence of lines emanating from outward skewed points in each region.  Scalebar 100$\mu$m.  B) Freely suspended droplet under cross polarizers pitch 2$\mu$m. The Robinson \cite{ROBINSON} texture with a +2 disclination line in the cholesteric structure. Scalebar 60$\mu$m.}
\end{figure}
  \section*{Results} 
\begin{figure}[t]
 \centering
 \begin{tabular}{ c }
  \includegraphics[width=0.35\textwidth ]{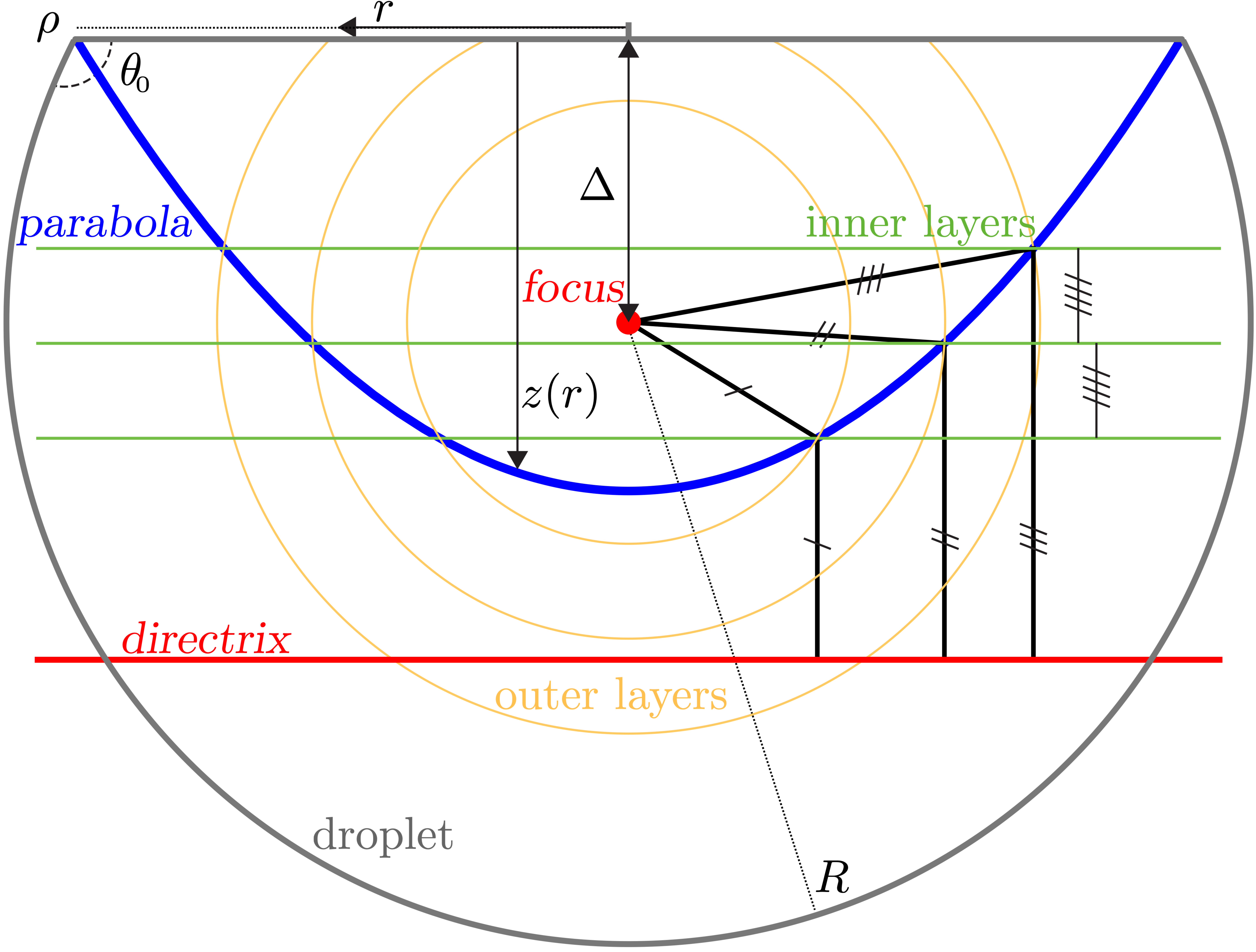}\\(A)\\
	\includegraphics[width=0.3\textwidth ]{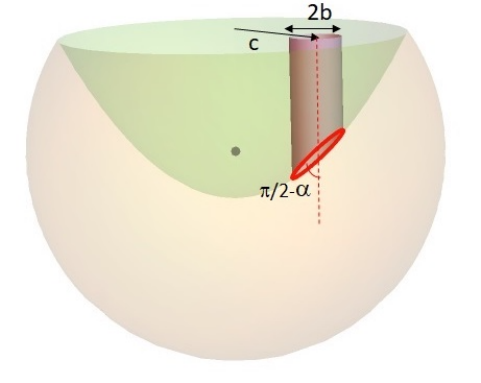}\\ (B)\\
	\end{tabular}
 \caption{   \label{fig:schematic} Schematic of the droplet.  {(A) A cross section by a plane perpendicular to the glass capillary substrate passing through the droplet center for a droplet containing a parabolic interface that defines a tilt grain boundary at the intersection of the systems of equidistant planes parallel to the droplet-substrate interface with concentric spheres centered at the droplet center.} At the paraboloid, the discontinuity in the layer normals scatters light in bright field. {(B)} {The 3D sketch of the parabolic tilt grain boundary with the varying disorientation angle $\alpha$ at the intersection of the equidistant parallel flat and spherical layers, above and below the paraboloid, respectively. The projection of a circle in the $xy$-plane with radius $b$ and centered at a distance $c$ from the centerline of the droplet is  an ellipse with minor axis $b$ and major axis $b\sqrt{1+4A^2c^2}$.  It corresponds to the intersection of the paraboloid with a tilted plane with tilt angle $\alpha$. The ellipse eccentricity is $e=\sin\alpha=\frac{2Ac}{\sqrt{1+4A^2c^2}} $.} } 
 \end{figure}

 We create cholesteric droplets in glycerol and, because they are less dense than the solvent they slowly cream under gravity.  The anchoring conditions between the mesogen and either the glass or glycerol are planar, requiring the pitch axis to be perpendicular to the interfaces.
When the glass is carefully cleaned the droplets do not wet the glass and remain spherical even on contact; however, when the glass has only been cleaned with isopropanol, the droplets stick to the surface and are deformed with a contact angle close to 124$^o$.  We can see in Fig. \ref{fig:Droplet24pc_2} that a parabolic interface separates two regions.  Even without direct observation of the cholesteric stripes through cross-polarizers as in Fig. \ref{fig:Droplet24pc}B, this demonstrates the layered nature of the cholesteric indirectly, {in the same spirit in which George Friedel deduced the layered structure of smectics by observing the FCD texture of confocal} {ellipses} {and hyperbola under the optical microscope}.  Spherical and flat layers join continuously across an interface if and only if it is a paraboloid, whose focus is the center of the sphere: indeed, a standard definition of a parabola is the locus of points that are equidistant from the focus and the directrix as sketched in Fig. \ref{fig:schematic}A.  {The congruent lines demonstrate the definition of the parabola: the locus of points (in blue) that are equidistant from the focus and the directrix (both in red).  It follows by construction that the outer, equally-spaced spheres (yellow) come into registry with the inner, equally-spaced planes (green).  With degenerate planar anchoring, these pseudolayers satisfy the boundary conditions and equal spacing.}
 Following the diagram, let $R$ be the radius of the droplet, $\rho$ be the radius of the droplet's disc-like interface with the glass, $\theta_0$ be the contact angle, $\Delta$ be the distance from the parabola focus to the slide.  Then, applying the Pythagorean theorem, we find the height of the parabola $z(r)$ as a function of the radius in cylindrical coordinates:   

\begin{equation} \nonumber
z(r)=\frac{{r^2-\rho^2}}{2\left(\sqrt{\Delta^2+\rho^2}-\Delta\right)}
\end{equation}
Since $\Delta=R\cos(\pi-\theta_0)$ and $\rho=R\sin(\pi-\theta_0)$, we find that the bottom of the paraboloid is $z(0)=\frac{1}{2}R(1-\cos\theta_0)\approx 0.82R$ which agrees extremely well with the observations.  

 Further inspection of the interface reveals the presence of curves on the paraboloid that, when viewed from below are nearly circular with lines emanating from foci, in the manner of FCDs (Figs. \ref{fig:Droplet24pc_2} and \ref{fig:Droplet24pc}A).  From the side they appear to be elliptical.  Note that for a general paraboloid in cylindrical coordinates, $z=A(r^2-\rho^2)$ and the intersection of the paraboloid with the plane $z=\delta-A\rho^2+x\tan\alpha$ (shifted along $z$ by $\delta$ from the bottom of the paraboloid and at an angle $\alpha$ with the $xy$-plane) projects to the circle $A(x-A^{-1}\frac{1}{2}\tan\alpha)^2 + Ay^2-\delta-\frac{1}{4A}\tan^2\alpha=0$ in the $xy$-plane (whenever $\tan^2\alpha + 4\delta\ge0$) {(Fig. \ref{fig:schematic}B)}. To be concrete, if the circle in the $xy$-plane has radius $b$ and is centered at a distance $c$ from the centerline of the droplet then the ellipse has minor axis $b$ and major axis $b\sqrt{1+4A^2c^2}$.  On the tilted plane this is an ellipse with eccentricity $e=\sin\alpha$  {Note that when $e=c=0$ the ellipse reduces to a circle centered on the paraboloid axis, while the hyperbola reduces to a straight line passing through the droplet center.  } We are led to conclude that the closed curves on the parabola are, indeed, ellipses and are the telltale signatures of focal conic domains.  Indeed, in both the bottom-up (Fig. \ref{fig:Droplet24pc}A) and side view (Fig. \ref{fig:Droplet24pc_2}) we not only observe the ellipses but we also observe lines perpendicular to them that we will show are the hyperbol\ae\ that are confocal with the ellipses of the FCDs.  

\begin{figure}
 \centering  \includegraphics[width=0.44\textwidth ]{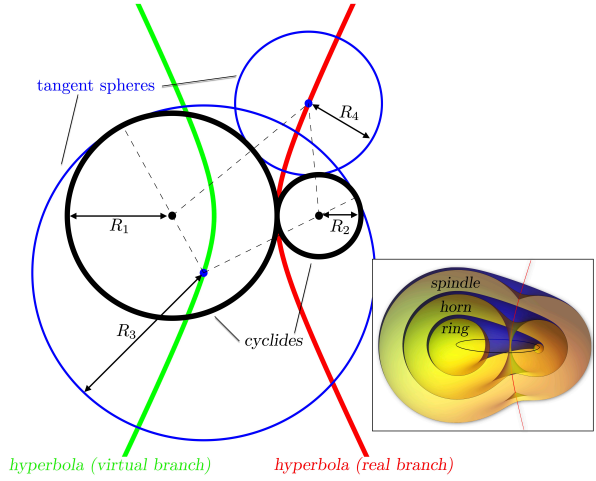} 
 \caption{ \label{fig:schematic2} A cross section of a horned cyclide of Dupin, in the plane of the hyperbola.  The centers of each black circle go through the ellipse perpendicular to the page (not shown).  The real branch of the hyperbola (red) traces out the cusps on concentric spindle cyclides and threads the hole of the concentric ring cyclides.  Noting that $(R_4+R_1)-(R_4+R_2)=R_1-R_2$ is constant, we see that this branch indeed sweeps out a hyperbola and, because the small and large black circles have radii that are collinear with radii of the upper blue circle, a family of equally spaced cyclides can be constructed with circles tangent to the negative Gaussian curvature regions of the cyclide.  In the text we prove that this construction works for the entire three-dimensional cyclide.  Similarly, the virtual branch (green) is a hyperbola because $(R_3-R_1) - (R_3-R_2)=R_1-R_2$ is also constant.  Spheres centered on this branch are tangent to the positive Gaussian curvature regions of the cyclide.  Inset shows the three types of cyclides, spindle, horn, and ring with red hyperbola and black ellipse.}  \end{figure}

Recall that FCDs have remarkable geometric properties in relation to spheres.  As prominently noted by Kleman and Sethna \cite{KlemanSethna}, it is possible to excise a cone from a collection of equally-spaced, concentric spheres (with the cone apex coincident with the common center) and replace it with equally-spaced sections of cyclides with {\sl no} discontinuity in the layer normal field.  As a result, there is no discontinuity in the local dielectric complexion and the interfaces between the FCD and spherical regions is not observed.  In Fig. \ref{fig:schematic2} we demonstrate the geometry of the cyclides and the associated hyperbola in the plane containing the hyperbola.  The black circles are the largest and smallest meridional cycles on the cyclide while the blue circles are cross sections of spheres.  How can we demonstrate that the tangency conditions shown in this plane extend to the remaining part of space, absent cylindrical symmetry?  Recall that equally-spaced layers have a hidden Poincar\'e symmetry \cite{AMCK} that can be used, for instance, to Lorentz transform a general cyclide to a symmetric torus.  Under the same Lorentz transformation, spheres map to spheres.  In the symmetric torus case, cylindrical symmetry insures that the blue circle tangent to the two black circles becomes a blue sphere (with center on the axis of symmetry) tangent to a black torus.  Lorentz transforms preserve intersections and tangencies and it follows that the general cyclide is tangent along a circle to a sphere centered on the hyperbola, the transform of the axis of symmetry.  Note that in the limit that the blue sphere center moves off to infinity along real branch, it becomes a plane tangent to the cyclide and the cone becomes an infinite cylinder.  This construction is the basis for the construction of large angle grain boundaries \cite{Kleman2000} and can also give rise to spatially varying eccentricity in thick enough freestanding films \cite{Y1,Y2,Y3,Y4}
or by confining a hybrid-aligned smectic with curved interfaces \cite{BellerFlower}  by manipulating interface behavior
with colloidal particles. Note that a sphere centered on the virtual branch at the diametrically opposed infinity {\sl also} is a plane tangent to the cyclide.  Bringing that center in from infinity, we see that now there are spheres tangent to the positive curvature regions of the cyclide.  
 
With this in hand, we propose the following structure for the cholesteric layers in the deformed droplet: inside the paraboloid the ellipses cap off circular cylinders that lie along $\hat z$ and are filled with concentric, nonpositive Gaussian curvature regions of cyclides.  The angle between the plane of the ellipse and the normal to the layers is $\alpha$, as above.  In the region outside the paraboloid, right circular cones centered on the focus of the paraboloid intersect the ellipses and in them we replace the spherical layers of Fig. \ref{fig:schematic} with concentric, equally-spaced pieces of cyclides.  The ray emanating from the center of the cone to the droplet surface is at an angle $\alpha$ to the plane of the ellipse and so the ellipse renders a rotation of the layer normals by $2\alpha$ \cite{Kleman2000}. Note that in this construction both negative {\sl and} positive Gaussian curvature regions contribute. This construction explains the ``inverse flower'' textures observed in \cite{meyer,BellerFlower} and here where the hyperbol\ae\ go through the outer foci of the ellipses, contrary to the hallowed law of corresponding cones.  Fortunately, nothing is lost.  We extend the law of corresponding cones to allow for the virtual branches of the hyperbola to meet at one point -- here at the focus of the paraboloid or, equivalently, at the center of the original spherical droplet.  Moreover, to the best of our knowledge this is the first report on the observation of an FCD domain in which the layers change the sign of their Gaussian curvature.

\begin{figure}[ht]
 \centering
\begin{tabular}{ c c  }
\begin{minipage}{0.35\textwidth}
  \includegraphics[width=\textwidth,angle=180]{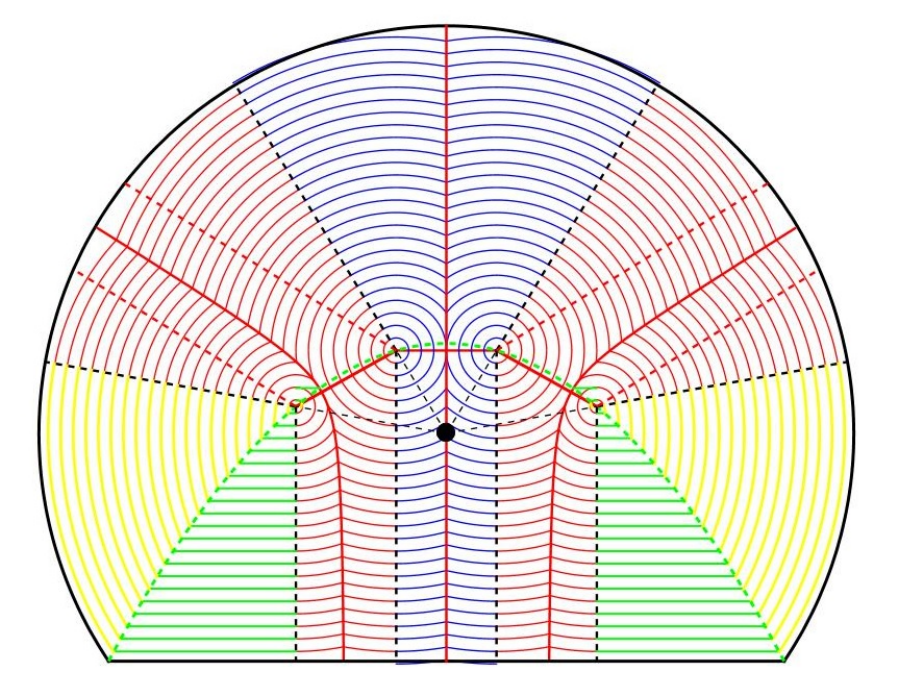}
  \end{minipage}&
  \begin{minipage}{0.10\textwidth}
 \includegraphics[width=\textwidth,angle=180]{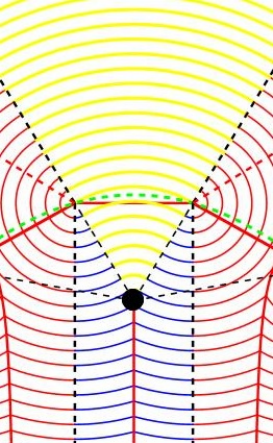}
 \end{minipage}\\(A)&(B)\\
 \end{tabular}
\caption{{Sketch of the proposed internal structure of the droplet in terms of cholesteric layers in the cross-section, which is perpendicular to the substrate and contains the droplet center.} {The disclinations are depicted by solid lines. Dashed lines show the boundaries between domains.There is no discontinuity at the boundaries of FCDs and they are smoothly embedded into the systems of mother flat and spherical layers. The dashed green line corresponds to the paraboloid. }{ We propose two different models that differ only by the structure of the central part. In (A) the TFCD fills the space below as well as above the horizontal equatorial plane and consequently the straight line disclination passing through the droplet center spans the droplet. In (B) the structure is the same as in (A) except between the horizontal equatorial plane and the interface with glycerol the space is filled by the mother concentric spherical layers and ,consequently, the straight line disclination of the TFCD only spans half the droplet. Such an alternate structure could be preferable as it shortens the central disclination.}
\label{fig:Sketch1}
} 
\end{figure}

\begin{figure}[t]
 \centering
\begin{tabular}{ c c  }

\begin{minipage}{0.28\textwidth}
\includegraphics[width=1\textwidth,angle =180]{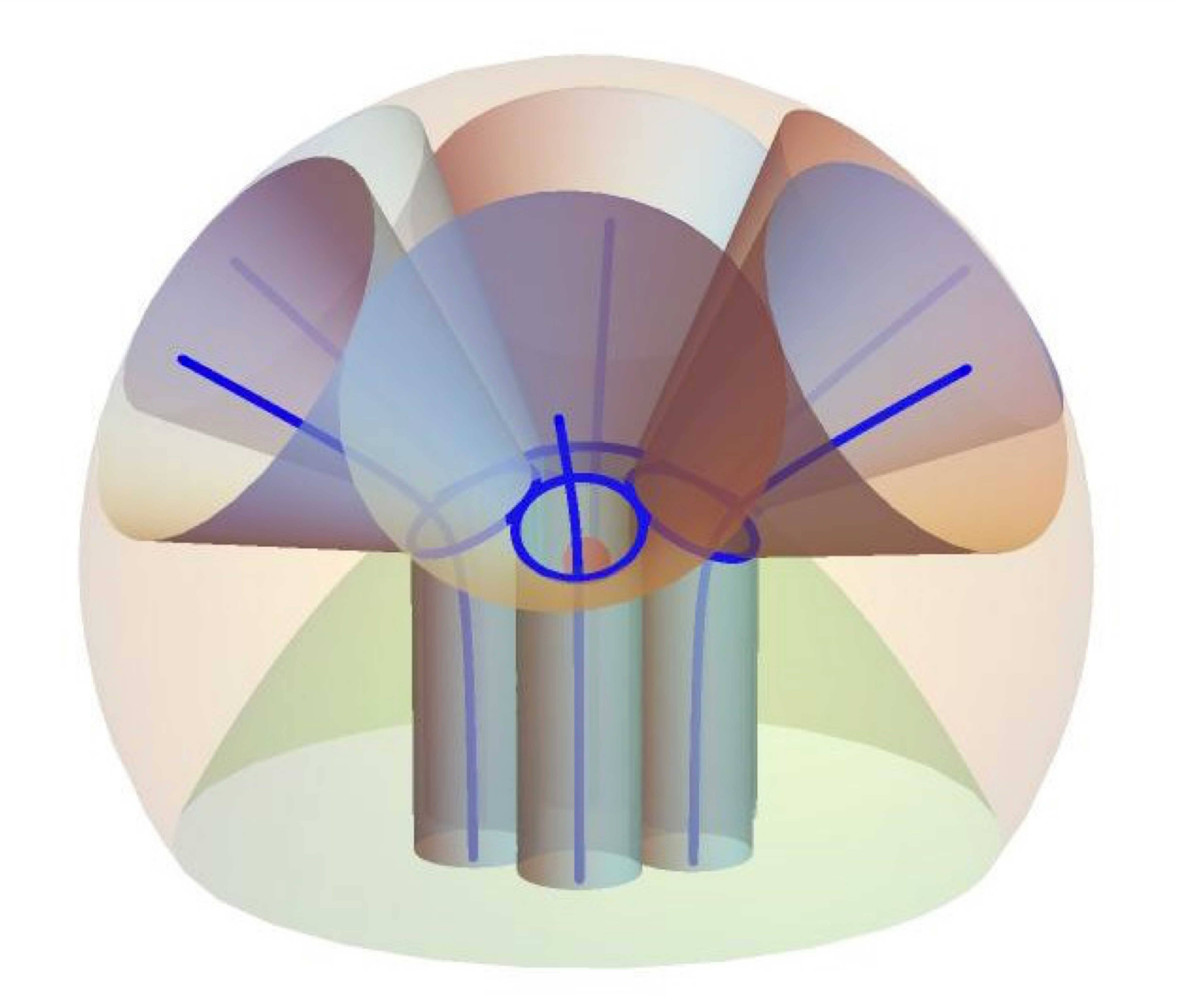} 
\end{minipage}&
\begin{minipage}{0.18\textwidth}
\includegraphics[width=1\textwidth]{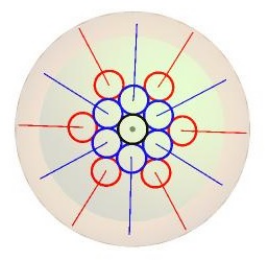} 
\end{minipage}\\

(A)&(B)\\
\end{tabular}
 \caption{(A) Three-dimensional structure of focal conic domains shown in cross section in Fig. \ref{fig:Sketch1}.  The cylinders and cones contain concentric cyclides; the remaining volume is filled with flat layers inside the paraboloid and spherical layers outside the paraboloid.  (B) The construction can be made hierarchically in the spirit of Apollonian packings.  The ellipses fill the paraboloid just as their circular projections onto the plane of the glass slide fill the plane. {The central FCD is a TFCD with its ellipse and hyperbola reduced to a circle and straight line. The central TFCD can be recognized in the experimental Fig. 9.}  }
\label{fig:Sketch2}
\end{figure}

{The resulting internal structure of the droplet in terms of cholesteric layers in a cross-section perpendicular to the substrate and containing the center is depicted in Fig. \ref{fig:Sketch1}. The mother spherical and flat layers in some places are replaced by FCDs. Blue corresponds to layers of the central toric FCD (TFCD), whose ellipse and hyperbola reduced, respectively, to the circle and straight line passing through the circle center. Red depicts layers of two other FCDs on both sides of the TFCD. Mother spherical layers concentric around the droplet center are shown in yellow, and mother flat layers are in green. Note that in all three FCDs, the sign of Gaussian curvature of the layers switches from negative to positive within each FCD. Namely, in the central domain,
blue layers have negative Gaussian curvature inside the cylinder but have positive Gaussian curvature outside the cylinder yet remain sections of Dupin cyclides (nested tori in this case). In the left and right domains, the layers in red have negative curvature inside the cylinders and regions of positive curvature outside the cylinders depicted by the dashed red lines. The sketch of the corresponding 3D structure is depicted in Fig. \ref{fig:Sketch2}A.} The remaining  volume of the droplet can again be broken into analagous cylinder/cone FCDs creating the hierarchical bouquet in Fig. \ref{fig:Sketch2}B.  Each ``flower'' demonstrates the energetic balance between two configurations: 1) the flat/sphere configuration has lower curvature energy than the cyclide sections but has a director singularity on the entire two-dimensional interface between the inner and outer regions; and 2) the cyclide configuration, with higher curvature energy, reduces the singular set to a one-dimensional set of curves, the ellipse and the hyperbola.  As the angle of the director discontinuity decreases towards the top of the adsorbed droplet, the associated interfacial energy likewise decreases and, at some point, it is no longer economical to replace the flat/sphere regions with focal domains.  

Not only does this model reproduce the broad geometrical structure of the droplet, but it provides, as well, numerical predictions.  In addition to predicting the height of the parabola in terms of the contact angle, the geometric construction relates, for instance, the radius ($b$) and center of each circle ($c$) in the $xy$-plane (Fig. \ref{fig:Droplet24pc}A) to the height of the center of the ellipse ($h$), shown in Fig. \ref{fig:Droplet24pc_2}.  Noting the equation for the paraboloid, we find $h=(z_++z_-)/2$ with $z_{\pm}=z(c\pm b)$.  In the bottom view of the sample we have a central circle, a primary ring of circles, and a secondary ring of circles.  In the primary ring the average 
values of $b$ and $c$ are $7.4\mu$m and $21\mu$m, while in the outer ring they are, on average, $6.4\mu$m and $33\mu$m, respectively.  Using $R=65\mu$m, $\Delta=36\mu$m, and $\rho=54\mu$m, we predict $h_{\rm in}= 42\mu$m and $h_{\rm out}=31\mu$m.  Measuring (with significant noise) the side image, we find $\overline{h}_{\rm in}=47\mu$m and $\overline{h}_{\rm out}=34\mu$m.  Likewise, we can calculate the expected angle between the asymptotes of the outer hyperbol\ae, $2\alpha$, via $\csc^2\alpha = \sqrt{1+4b^2/(z_+-z_-)^2}$ and find $2\alpha\approx 97^\circ$ compared to the measured $2\overline{\alpha}\gtrsim 85^\circ$.  Since it is difficult to identify the circles in the bottom view with the ellipses in the side view, we cannot, at this time, offer more precise numbers.   Suffice to say, the agreement between the model and measurement is encouraging if not convincing.

 \begin{figure}[h]
 \centering
  \includegraphics[width=0.44\textwidth ]{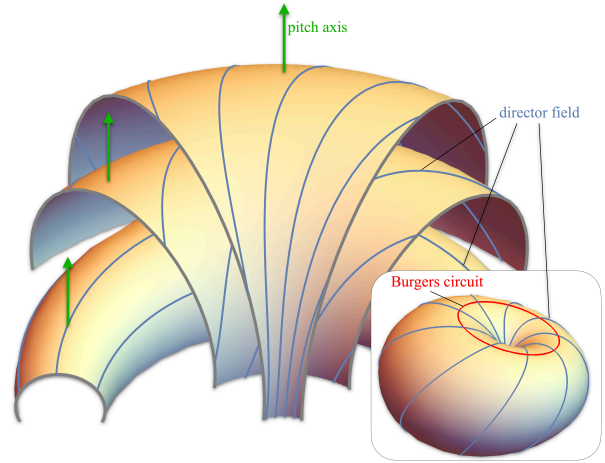} 
 \caption{{Configuration of the director field on cyclides of Dupin. Within any FCD, on each cyclide, the director rotates around the hyperbola, including the central toric FCD, for which the hyperbola reduces to a straight line. Inset: Measuring the director field along the Burgers circuit (red) on a spindle cyclide. The angle between the red curve and the director is constant and thus the director rotates by $2\pi$ along the circuit. Since the surface no longer has a handle, this implies that there is a defect at the cusp.}  }
 \label{fig:Sketch21}
\end{figure}

Contrary to smectic A textures, the normal  to the layers is not the only parameter to describe the textures: the director field which is perpendicular to the layer normal has to be defined. This clearly increases the complexity of the texture.  Fortunately, the geometry here allows us to make some progress.  When the integral curves of the local pitch axis are straight lines, there is no obstruction to adding a director field everywhere perpendicular to the pitch axis that rotates smoothly \cite{cholgeom}. Since all the geometries we have used consist of equally-spaced, parallel layers, this condition is satisfied and we can build a cholesteric-like winding on the equally-spaced cyclides shown in Fig. \ref{fig:Sketch21}.  However, topological considerations create a global obstruction in some cases.  Consider, for instance, the spherical droplet shown in Fig. \ref{fig:Droplet24pc}B.  Were this a smectic A droplet, it could be filled with concentric spherical layers parallel to the boundary.    However, if we try to decorate each spherical pesudolayer with a director field, perpendicular to the normal, we cannot.  The Poincar\'e-Brouwer-Hopf theorem requires defects with total winding of $4\pi$.  Indeed, the Robinson texture shown in Fig. \ref{fig:Droplet24pc}B demonstrates this \cite{ROBINSON}.  The singular line running from the center to the surface is required by topology and captures a winding of $4\pi$.  In a droplet with no focal conic domains, as described in Fig. \ref{fig:Sketch1}, this is the only necessary singularity and will still be required even when the spherical pseudolayers of the cholesteric are deformed to the squashed droplet shape.  The introduction of the focal conic domains adds to the complexity of the cholesteric.  When the domains are built from ring cyclides there is no problem -- ring cyclides have the topology of a torus and, as a result, the director field need not have any defects.  Eventually, however, the holes pinch off and we must build with sections of spindle cyclides (see inset of Fig. \ref{fig:schematic2}).  When this happens, the director texture will be required to have a $+1$ winding defect at the cusp, leading to a line of defects along the confocal hyperbola.  These lines can be seen emanating from the far foci of the ellipses in the bottom view of the sample.  To see this, consider the spindle cyclide shown in the inset of Fig. \ref{fig:Sketch21}.    The director field makes a constant angle with the meridians of the cyclide (this angle rotates smoothly along the pitch axes).  The Burgers circuit therefore measures a winding of $+2\pi$. 

{The side-view of the droplet (Fig. 2) does not show any evidence of a disclination with winding number 2 either spanning the diameter of the droplet or even ending at the center.
The dot in the center in Figs. 3A and 9 might be either a projection of the disclination to which a hyperbola reduces when an FCD reduces to a TFCD. Fig. 9 shows the presence of a circle defect centered at the droplet center in the central part of the droplet, thus suggesting that the dot in the center is a projection of the straight line of the TFCD. However, which of the scenarios realized is subject to further experimental verification.}

\begin{figure}
 \centering
  \includegraphics[width=0.44\textwidth ]{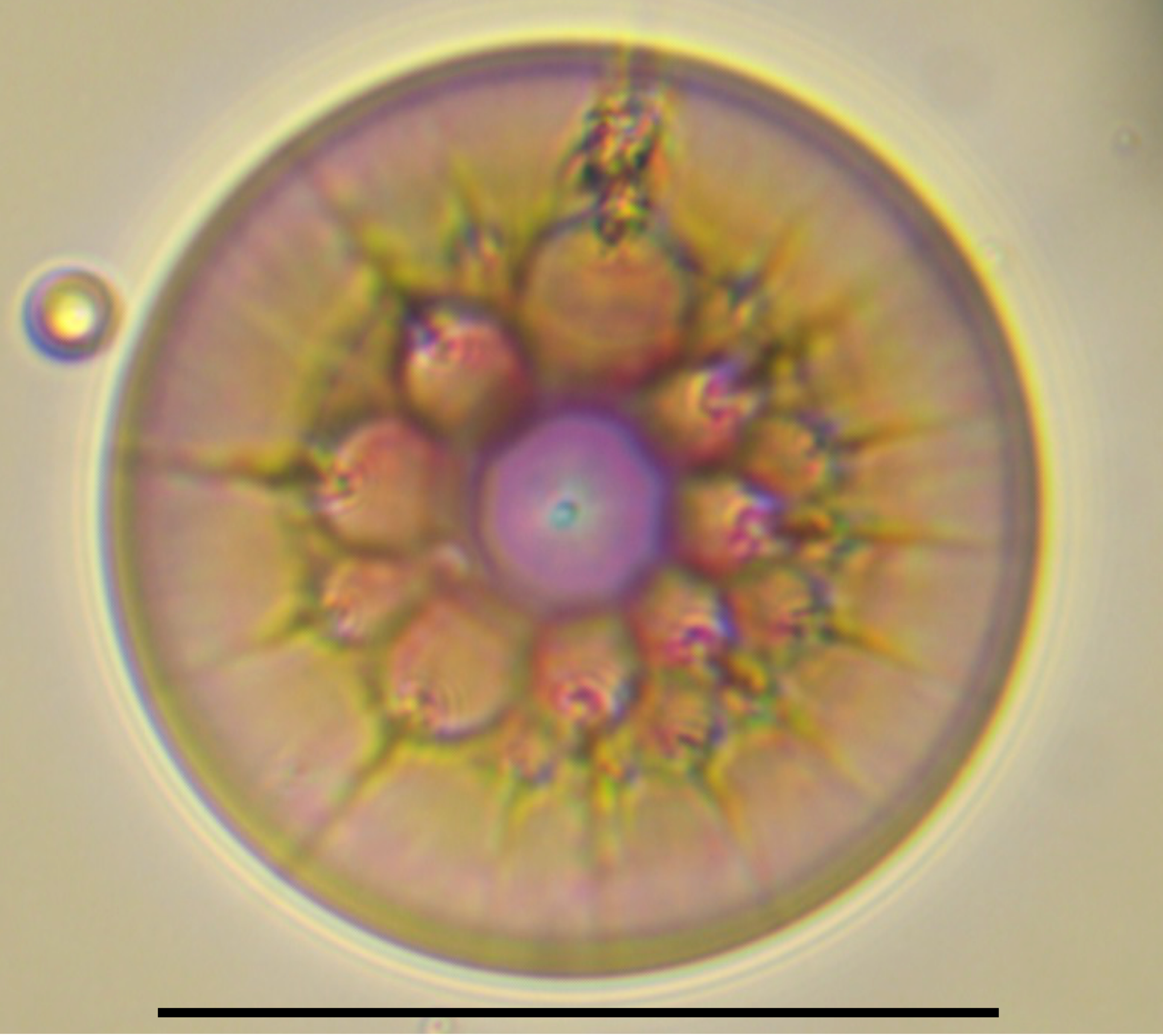} 
 \caption{The hyperbol\ae\ can be observed to wobble and divide, in some cases, into pairs of disclination lines.  The rings seen around the foci indicate the complex dielectric structure resulting from a complex disclination core in the cholesteric. {Note that the circular defect line belonging to the central  TFCD is schematically depicted in Fig. 7B. The dot in the center of the circular defect is a projection of the straight line defect to which the hyperbola reduces in a TFCD.} Scalebar is100$\mu$m, pitch is $0.35\mu$m. }
 \label{fig:twist}
\end{figure}

Of course if there were exactly one cylinder/cone FCD, this would account for the total required charge of $4\pi$ -- one defect line emanating from the parabolic focus ending on the spherical part of the droplet and a second $+1$ line, also emanating from the focus, ending on the flat part of the boundary. This is not the case: the bottom view of the droplet shows many domains.   Note that $+1$ disclination lines can ``escape into the third dimension'' in pure nematics \cite{rmeyer}, in a cholesteric this would create a defect in the pitch axis.  For instance in a disc of radius $\pi/2$, the nematic texture (in cylindrical coordinates) ${\bf n}(r)=\hat z\cos r + \hat\theta\sin r$ has winding 
$+2\pi$ at $r=\pi/2$ and no winding at $r=0$.   However, at the origin there is not a unique pitch axis around which the director rotates.  We have only exchanged a $\chi$-disclination line for a $\lambda$-disclination line.  Because of this, the core of a $+1$ disclination line is not only larger than the molecular scale, but it can also show additional structure \cite{GarethPRL}.  Indeed, as shown in Fig. \ref{fig:twist}, we observe the hyperbol\ae\ splitting and twisting, presumably into a pair of disclination lines tracing out charge $-1/2$ defects in the pseudolayers as in Fig. \ref{fig:twist}.

If each FCD generates $2\pi$ winding in the director that lies in each pseudolayers, what happens to the topological constraint of total winding $4\pi$ in the textures we observe?  Note that in a packing of circles or ellipses, typically three of them are mutually tangent.  Since the director is continuous on these measuring circuits, it follows that  in the interstitials {\sl between} three tangent circles, there must be a winding of $-\pi$, charge $-1/2$.   Isolated charge $-1/2$ disclinations in the pseudolayers have no means of escape or reconfiguration and we expect their cores to be on the molecular scale.  Topological constraints are incorruptible and so we attribute our inability to observe these defect lines to their sub-optical size (no bright field scattering) and the optical turbidity of the droplet (not visible under cross polarizers).   Arduous freeze-fracture studies might be possible in the future.  

We also note that, though common, the texture we find and describe here is not the only observed structure.  When the droplet is smaller than $80\mu$m in diameter, we only observe a single row of flowers, as in Fig. \ref{fig:Sketch2}A.  Even smaller droplets with diameter less than $40\mu$m exhibit no FCD cylinder/cone textures.  Similarly, when we increase the pitch of the cholesteric, only larger droplets form the bouquet, suggesting that it is only the dimensional ratio of pitch to diameter that determines the equilibrium structure.
{We noticed that the history of the sample is crucial for the formation of focal conic flowers. Usually, a bouquet of FCDs similar to those shown in Figs. 2 and 9 forms sometime (several days or weeks) after a freely suspended droplet with the Robinson texture (Fig. 3) attaches to the glass substrate. Liquid crystal material stuck to the substrate during the filling of the capillary might form a sessile droplet, but its texture often is cluttered, far from that of an FCD bouquet. Once such a sessile droplet with a cluttered texture is formed, heating it to the isotropic phase does not result in the formation of a FCD bouquet because of the strong anchoring of liquid crystal molecules at the glass substrate.}
Analogous textures appear in Janus particle systems of smectic A but with much lower regularity \cite{Yodh_D1SM01623G}.  With our model in mind, it would be interesting to return to that system to search for purely smectic bouquets.

The reader might notice that no energetic considerations were made in our construction.  That the system is, indeed, finding a ground state is not only implied by the persistence of the droplet texture but also by noting that the geometry alone does not set the size of the cylinder/cone flowers.  The fact that they are relatively uniform in radius strongly suggests that they are minimizing an energy and not the result of kinetic trapping which would, as in typical FCD textures, present as a large assortment of sizes.  Apollonius conceived of conics over two millennia ago only to now reveal themselves so perfectly in a state of matter with uniform density.  What's next?

\section*{\bf Materials and Methods}

 The cholesteric liquid crystal (ChLC) phase was prepared by mixing a nematic phase E3100-100 (from Merck liquid crystals) with the chiral agent S811 (from XARLM, China).   The optical indices measured in the nematic phases E3100-100  are   $n_o=1.57$ (o=ordinary) and $n_e=1.72$ (e=extraordinary) and thus the birefringence is $\Delta n=0.15$.
 Taking into account that the typically for the given chiral dopant HTP=12.77 $\mu{m}^{-1}$, the mixtures of the concentrations from 4 to 24 weight percent were prepared in order to obtain cholesteric between 0.33$\mu$m and 2$\mu$m.  Suspensions of cholesteric droplets were prepared stirring the LC material in glycerol.  After waiting a few minutes, the emulsion was gently sucked into a  glass capillary (VitroCom, USA) -- either a  0.4mm $\times$ 4mm  flat rectangular capillary or  a 1 mm square capillary.  The dispersity of the droplet diameter  depends on the stirring conditions and varies from 50 to 150 $\mu$m. The capillary was then  kept horizontal. After some time, the droplets move to the capillary surface and stick to the glass surface. The droplet structure was followed by optical microscopy (Microscope MEIJI MT9930L, Camera DeltaPix 12MP EXMOR) over several weeks.

 Glycerol is known to promote degenerate planar anchoring of the director, which implies homeotropic anchoring of the pitch axis (the twist direction). With such anchoring, the droplets suspended in glycerol exhibit a Robinson or onion structure  where the pitch direction is radial \ref{fig:Droplet24pc}B. Quasi-spherical parallel equidistant cholesteric pseudolayers self-assemble in droplets. When the cholesteric liquid crystal is doped with a dye, this microdroplet geometry has been  used to promote  a tunable and omnidirectional microlaser  \cite{Humar2010}.The director field necessarily contains defects, called disclinations, because of topological constraints due to the spherical geometry. Spherical symmetry of the cholesteric layers in such a droplet is broken by one (or two) radial disclination(s) of total topological strength +2 (+1 each). This (these) line disclination(s) is (are) required by the Poincar\'e-Hopf-Brouwer theorem since the director lies tangent to the quasi-spherical pseudolayers.

\begin{acknowledgments}
 We thank Ivan Dozov (LPS, Paris-Saclay) for providing the nematic liquid crystal material and giving us pertinent advice on microscopy observations. We acknowledge M\'elanie Lebental  (C2N, Paris-Saclay) who brought our attention on the role of the glass surface on the flower architecture of the droplets. Yu.N. thanks the Ministry of Education and Science of Ukraine and the Ministry of Defence of Ukraine for support.  Yu. N. is grateful to ISMO, LPS and ENS Paris-Saclay for hospitality, with a special thanks to Q. Kou (ISMO). Y. N. received financial support
from Fondation de Coop\'eration Scientifique ``Campus
Paris-Saclay'' through the MILACHOL3D project of RTRA
( No P2013-0561T) and from the Erasmus Mundus
program Monabiphot. RDK thanks the Institute for Theoretical Physics at Utrecht University for their hospitality when this work was initiated.  RDK was supported by a Simons Investigator Grant from the Simons Foundation. 
 \end{acknowledgments}
 


\begin{thebibliography}{29}
\providecommand{\natexlab}[1]{#1}
\providecommand{\url}[1]{\texttt{#1}}
\expandafter\ifx\csname urlstyle\endcsname\relax
  \providecommand{\doi}[1]{doi: #1}\else
  \providecommand{\doi}{doi: \begingroup \urlstyle{rm}\Url}\fi

\bibitem[Planer(1861)]{Planer1861}
Julius Planer.
\newblock Notis \"uber das cholestearin.
\newblock \emph{Annalen der Chemie und Pharmacie}, 118:\penalty0 25--27, 1861.

\bibitem[Nazarenko et~al.(2018)Nazarenko, Kurik, Klimusheva, Gotra, Sorokin,
  and Lisetski]{NAZARENKO201829}
V.~Nazarenko, M.V. Kurik, G.V. Klimusheva, Z.Yu. Gotra, V.M. Sorokin, and L.M.
  Lisetski.
\newblock Liquid crystals in ukraine and ukrainians in liquid crystals.
\newblock \emph{Journal of Molecular Liquids}, 267:\penalty0 29--33, 2018.
\newblock ISSN 0167-7322.
\newblock \doi{https://doi.org/10.1016/j.molliq.2018.01.053}.
\newblock URL
  \url{https://www.sciencedirect.com/science/article/pii/S0167732217353382}.
\newblock Special Issue Dedicated to the Memory of Professor Y. Reznikov.

\bibitem[Reinitzer(1888)]{Reinitzer}
F.~Reinitzer.
\newblock Beitr\"age zur kenntnis des cholesterins.
\newblock \emph{Monatshefte f\"ur Chemie}, 9, 1888.

\bibitem[Gennes(1993)]{dgp}
Pierre-Gilles~de. Gennes.
\newblock \emph{The physics of liquid crystals}.
\newblock International series of monographs on physics ; 83. Clarendon Press,
  Oxford, 2nd ed. edition, 1993.
\newblock ISBN 0198520247.

\bibitem[Dupin(1822)]{Dupin}
Charles Dupin.
\newblock \emph{Applications de G\'eom\'etrie et de M\'echanique, a la Marine,
  aux Ponts et Chauss\'ees, etc.}
\newblock Bachelier, 1822.

\bibitem[Grandjean and Friedel(1910)]{Friedel}
Fran\c{c}ois Grandjean and Georges Friedel.
\newblock Observations g\'eom\'eriques sur les liquides \`a coniques focales.
\newblock \emph{Bulletin de Minéralogie}, 33\penalty0 (8):\penalty0 409--465,
  1910.
\newblock \doi{10.3406/bulmi.1910.3454}.
\newblock URL
  \url{https://www.persee.fr/doc/bulmi_0366-3248_1910_num_33_8_3454}.

\bibitem[Canham(1970)]{Canham}
P.B. Canham.
\newblock The minimum energy of bending as a possible explanation of the
  biconcave shape of the human red blood cell.
\newblock \emph{Journal of Theoretical Biology}, 26\penalty0 (1):\penalty0
  61--81, 1970.
\newblock ISSN 0022-5193.
\newblock \doi{https://doi.org/10.1016/S0022-5193(70)80032-7}.
\newblock URL
  \url{https://www.sciencedirect.com/science/article/pii/S0022519370800327}.

\bibitem[Helfrich(1973)]{Helfrich}
W~Helfrich.
\newblock Elastic properties of lipid bilayers: Theory and possible
  experiments.
\newblock \emph{Zeitschrift f{\"u}r Naturforschung C}, 28\penalty0
  (11-12):\penalty0 693--703, 1973.
\newblock \doi{doi:10.1515/znc-1973-11-1209}.
\newblock URL \url{https://doi.org/10.1515/znc-1973-11-1209}.

\bibitem[{de Gennes}(1972)]{deGennes1972}
P.G. {de Gennes}.
\newblock An analogy between superconductors and smectics a.
\newblock \emph{Solid State Communications}, 10\penalty0 (9):\penalty0
  753--756, 1972.
\newblock ISSN 0038-1098.
\newblock \doi{https://doi.org/10.1016/0038-1098(72)90186-X}.
\newblock URL
  \url{https://www.sciencedirect.com/science/article/pii/003810987290186X}.

\bibitem[Aviles and Giga(1987)]{AvilesGiga}
P.~Aviles and Y.~Giga.
\newblock A mathematical problem related to the physical theory of liquid
  crystal configurations.
\newblock \emph{Proc. Centre Math. Appl.}, 1987:\penalty0 1--16, 1987.

\bibitem[Santangelo and Kamien(2003)]{bps}
C.~D. Santangelo and Randall~D. Kamien.
\newblock Bogomol'nyi, prasad, and sommerfield configurations in smectics.
\newblock \emph{Phys. Rev. Lett.}, 91:\penalty0 045506, Jul 2003.
\newblock \doi{10.1103/PhysRevLett.91.045506}.
\newblock URL \url{https://link.aps.org/doi/10.1103/PhysRevLett.91.045506}.

\bibitem[{J. B. Fournier} and {G. Durand}(1991)]{Fournier}
{J. B. Fournier} and {G. Durand}.
\newblock Focal conic faceting in smectic-a liquid crystals.
\newblock \emph{J. Phys. II France}, 1\penalty0 (7):\penalty0 845--870, 1991.
\newblock \doi{10.1051/jp2:1991113}.
\newblock URL \url{https://doi.org/10.1051/jp2:1991113}.

\bibitem[Kl\'eman and Lavrentovich(2009)]{Kleman}
M.~Kl\'eman and O.~D. Lavrentovich.
\newblock Liquids with conics.
\newblock \emph{Liquid Crystals}, 36\penalty0 (10-11):\penalty0 1085--1099,
  2009.
\newblock \doi{10.1080/02678290902814718}.
\newblock URL \url{https://doi.org/10.1080/02678290902814718}.

\bibitem[{Bouligand, Y.}(1972)]{Bouligand}
{Bouligand, Y.}
\newblock Recherches sur les textures des \'etats m\'esomorphes - 1. les
  arrangements focaux dans les smectiques : rappels et consid\'erations
  th\'eoriques.
\newblock \emph{J. Phys. France}, 33\penalty0 (5-6):\penalty0 525--547, 1972.
\newblock \doi{10.1051/jphys:01972003305-6052500}.
\newblock URL \url{https://doi.org/10.1051/jphys:01972003305-6052500}.

\bibitem[Sethna and Kl\'eman(1982)]{KlemanSethna}
James~P. Sethna and Maurice Kl\'eman.
\newblock Spheric domains in smectic liquid crystals.
\newblock \emph{Phys. Rev. A}, 26:\penalty0 3037--3040, Nov 1982.
\newblock \doi{10.1103/PhysRevA.26.3037}.
\newblock URL \url{https://link.aps.org/doi/10.1103/PhysRevA.26.3037}.

\bibitem[Kl{\'e}man and Lavrentovich(2000)]{Kleman2000}
M.~Kl{\'e}man and O.~D. Lavrentovich.
\newblock Grain boundaries and the law of corresponding cones in smectics.
\newblock \emph{The European Physical Journal E}, 2\penalty0 (1):\penalty0
  47--57, Apr 2000.
\newblock ISSN 1292-8941.
\newblock \doi{10.1007/s101890050039}.
\newblock URL \url{https://doi.org/10.1007/s101890050039}.

\bibitem[Beller et~al.(2013)Beller, Gharbi, Honglawan, Stebe, Yang, and
  Kamien]{BellerFlower}
Daniel~A. Beller, Mohamed~A. Gharbi, Apiradee Honglawan, Kathleen~J. Stebe, Shu
  Yang, and Randall~D. Kamien.
\newblock Focal conic flower textures at curved interfaces.
\newblock \emph{Phys. Rev. X}, 3:\penalty0 041026, Dec 2013.
\newblock \doi{10.1103/PhysRevX.3.041026}.
\newblock URL \url{https://link.aps.org/doi/10.1103/PhysRevX.3.041026}.

\bibitem[Robinson(1961)]{ROBINSON}
Conmar Robinson.
\newblock Liquid-crystalline structures in polypeptide solutions.
\newblock \emph{Tetrahedron}, 13\penalty0 (1):\penalty0 219--234, 1961.
\newblock ISSN 0040-4020.
\newblock \doi{https://doi.org/10.1016/S0040-4020(01)92215-X}.
\newblock URL
  \url{https://www.sciencedirect.com/science/article/pii/S004040200192215X}.

\bibitem[Alexander et~al.(2010)Alexander, Chen, Matsumoto, and Kamien]{AMCK}
Gareth~P. Alexander, Bryan Gin-ge Chen, Elisabetta~A. Matsumoto, and Randall~D.
  Kamien.
\newblock Power of the poincar\'e group: Elucidating the hidden symmetries in
  focal conic domains.
\newblock \emph{Phys. Rev. Lett.}, 104:\penalty0 257802, Jun 2010.
\newblock \doi{10.1103/PhysRevLett.104.257802}.
\newblock URL \url{https://link.aps.org/doi/10.1103/PhysRevLett.104.257802}.

\bibitem[Meyer et~al.(2007)Meyer, Nastishin, and Kleman]{Y1}
Claire Meyer, Yuriy Nastishin, and Maurice Kleman.
\newblock Kinked focal conic domains in a sma.
\newblock \emph{Molecular Crystals and Liquid Crystals}, 477\penalty0
  (1):\penalty0 43/[537]--53/[547], 2007.
\newblock \doi{10.1080/15421400701732449}.
\newblock URL
  \url{https://www.tandfonline.com/doi/abs/10.1080/15421400701732449}.

\bibitem[Kleman et~al.(2006)Kleman, Meyer, and Nastishin]{Y2}
M.~Kleman, C.~Meyer, and Yu.~A. Nastishin.
\newblock Imperfections in focal conic domains: the role of dislocations.
\newblock \emph{Philosophical Magazine}, 86\penalty0 (28):\penalty0 4439--4458,
  2006.
\newblock \doi{10.1080/14786430600724496}.
\newblock URL \url{https://doi.org/10.1080/14786430600724496}.

\bibitem[Nastishin et~al.(2008)Nastishin, Meyer, and Kleman]{Y3}
Yu.~A. Nastishin, C.~Meyer, and M.~Kleman.
\newblock Imperfect focal conic domains in a smectics: a textural analysis.
\newblock \emph{Liquid Crystals}, 35\penalty0 (5):\penalty0 609--624, 2008.
\newblock \doi{10.1080/02678290802041263}.
\newblock URL \url{https://doi.org/10.1080/02678290802041263}.

\bibitem[Nastishin and Meyer(2023)]{Y4}
Yuriy~A. Nastishin and Claire Meyer.
\newblock Imperfect defects in smectics a.
\newblock \emph{Liquid Crystals Reviews}, 0\penalty0 (ja):\penalty0 1--47,
  2023.
\newblock \doi{10.1080/21680396.2023.2181880}.
\newblock URL \url{https://doi.org/10.1080/21680396.2023.2181880}.

\bibitem[Meyer et~al.(2009)Meyer, Le~Cunff, Belloul, and Foyart]{meyer}
Claire Meyer, Loic Le~Cunff, Malika Belloul, and Guillaume Foyart.
\newblock Focal conic stacking in smectic a liquid crystals: Smectic flower and
  apollonius tiling.
\newblock \emph{Materials}, 2\penalty0 (2):\penalty0 499--513, 2009.
\newblock ISSN 1996-1944.
\newblock \doi{10.3390/ma2020499}.
\newblock URL \url{https://www.mdpi.com/1996-1944/2/2/499}.

\bibitem[Beller et~al.(2014)Beller, Machon, \ifmmode~\check{C}\else
  \v{C}\fi{}opar, Sussman, Alexander, Kamien, and Mosna]{cholgeom}
Daniel~A. Beller, Thomas Machon, Simon \ifmmode~\check{C}\else \v{C}\fi{}opar,
  Daniel~M. Sussman, Gareth~P. Alexander, Randall~D. Kamien, and Ricardo~A.
  Mosna.
\newblock Geometry of the cholesteric phase.
\newblock \emph{Phys. Rev. X}, 4:\penalty0 031050, Sep 2014.
\newblock \doi{10.1103/PhysRevX.4.031050}.
\newblock URL \url{https://link.aps.org/doi/10.1103/PhysRevX.4.031050}.

\bibitem[Meyer(1973)]{rmeyer}
Robert~B. Meyer.
\newblock On the existence of even indexed disclinations in nematic liquid
  crystals.
\newblock \emph{The Philosophical Magazine: A Journal of Theoretical
  Experimental and Applied Physics}, 27\penalty0 (2):\penalty0 405--424, 1973.
\newblock \doi{10.1080/14786437308227417}.
\newblock URL \url{https://doi.org/10.1080/14786437308227417}.

\bibitem[Pollard and Alexander(2023)]{GarethPRL}
Joseph Pollard and Gareth~P. Alexander.
\newblock Contact topology and the classification of disclination lines in
  cholesteric liquid crystals.
\newblock \emph{Phys. Rev. Lett.}, 130:\penalty0 228102, Jun 2023.
\newblock \doi{10.1103/PhysRevLett.130.228102}.
\newblock URL \url{https://link.aps.org/doi/10.1103/PhysRevLett.130.228102}.

\bibitem[Wei et~al.(2022)Wei, Jeong, Collings, and Yodh]{Yodh_D1SM01623G}
Wei-Shao Wei, Joonwoo Jeong, Peter~J. Collings, and A.~G. Yodh.
\newblock Focal conic flowers{,} dislocation rings{,} and undulation textures
  in smectic liquid crystal janus droplets.
\newblock \emph{Soft Matter}, 18:\penalty0 4360--4371, 2022.
\newblock \doi{10.1039/D1SM01623G}.
\newblock URL \url{http://dx.doi.org/10.1039/D1SM01623G}.

\bibitem[Humar and Mu\v{s}evi\v{c}(2010)]{Humar2010}
M.~Humar and I.~Mu\v{s}evi\v{c}.
\newblock 3d microlasers from self-assembled cholesteric liquid-crystal
  microdroplets.
\newblock \emph{Opt. Express}, 18\penalty0 (26):\penalty0 26995--27003, Dec
  2010.
\newblock \doi{10.1364/OE.18.026995}.
\newblock URL \url{http://opg.optica.org/oe/abstract.cfm?URI=oe-18-26-26995}.

\end{thebibliography}
\end{document}